\documentclass[aps,prl,twocolumn,nofootinbib]{revtex4-1}

\usepackage{amsmath,amssymb,bm}
\usepackage[dvips]{graphicx}
\usepackage{hyperref}
\usepackage{yfonts}
\usepackage{color}
\newcommand{\beq}{\begin{eqnarray}}
\newcommand{\eeq}{\end{eqnarray}}

\newcommand{\U}{\text{U}}
\renewcommand\d{\partial}
\begin{document}

\title{Magnetic monopoles and fermion number violation in chiral matter}

\author{Naoki Yamamoto}
\affiliation{Department of Physics, Keio University, Yokohama 223-8522, Japan}

\begin{abstract}
We show that the presence of a magnetic monopole in position space gives rise 
to a violation of the fermion number conservation in chiral matter. Using the 
chiral kinetic theory, we derive a model-independent expression of such a 
violation in nonequilibrium many-body systems of chiral fermions. 
In local thermal equilibrium at finite temperature and chemical potential, 
in particular, this violation is proportional to the chemical potential with a 
topologically quantized coefficient. These consequences are due to the 
interplay between the Dirac monopole in position space and the Berry monopole 
in momentum space. Our mechanism can be applied to study the roles of 
magnetic monopoles in the nonequilibrium evolution of the early Universe.
\end{abstract}
\maketitle

\emph{Introduction.}---% 
The magnetic monopole (in position space)%
\footnote{Note that we consider two different types of magnetic monopoles
in this paper: one in position space and the other in momentum space. 
To avoid possible confusion, we often call the former 
the Dirac monopoles \cite{Dirac:1931kp}
and the latter the Berry monopoles \cite{Berry}.}
is a hypothetical object that has close connections to various fundamental 
problems in theoretical physics: 
the quantization of electric charges \cite{Dirac:1931kp}, 
the grand unified theory \cite{tHooft:1974kcl,Polyakov:1974ek},
the proton decay \cite{Rubakov:1982fp,Callan:1982au},
evolution of the early Universe, and so on. Nonetheless, despite its 
relevance especially to the evolution of the early Universe, understanding 
of the roles of magnetic monopoles in nonequilibrium many-body systems 
remains elusive.

In this paper, we study the physical consequences of magnetic monopoles 
in nonequilibrium many-body systems of chiral (or Dirac) fermions.
Such an analysis becomes feasible thanks to the recent developments of 
the kinetic theory for chiral fermions, called the chiral kinetic theory \cite{Son:2012wh,Son:2012zy,Stephanov:2012ki,Chen:2012ca,Manuel:2013zaa,Manuel:2014dza,Chen:2014cla,Chen:2015gta,Hidaka:2016yjf,Hidaka:2017auj,Mueller:2017arw}, 
where the quantum corrections due to the chirality are incorporated as 
the Berry phase \cite{Berry} in momentum space.

We show that, when a magnetic monopole is present in chiral matter, 
it leads to a violation of the fermion number conservation. Although this is 
reminiscent of the Rubakov-Callan effect \cite{Rubakov:1982fp,Callan:1982au}, 
we note that our results are derived for nonequilibrium many-body systems 
and are model-independent. We also show that, when the system is in local 
thermal equilibrium characterized by a temperature $T$ and chemical potential $\mu$, 
the violation of the fermion number conservation is proportional to $\mu$ 
with a topologically quantized coefficient; see Eqs.~(\ref{main_equilibrium}) 
and (\ref{main_global2}) for the case of $\U(1)$ charges. As we will argue, 
these are the consequences of the interplay between two types of monopoles: 
the Dirac monopole in position space and the Berry monopole in momentum space.

Throughout the paper, we use the natural units $\hbar=c=1$.

\emph{Chiral kinetic theory.}---%
To make our paper self-contained, we will first review the chiral kinetic theory \cite{Son:2012wh,Son:2012zy,Stephanov:2012ki,Chen:2012ca,Manuel:2013zaa,Manuel:2014dza,Chen:2014cla,Chen:2015gta,Hidaka:2016yjf,Hidaka:2017auj,Mueller:2017arw}.
Although the chiral kinetic theory can be written in a Lorentz covariant manner,  
we will apply the chiral kinetic theory with a Berry curvature that do not have 
a manifest Lorentz covariance \cite{Son:2012wh,Son:2012zy,Stephanov:2012ki}, 
as it is simpler both conceptually and practically for our purpose. 
(This corresponds to a particular choice of the frame vector in the 
Lorentz-covariant chiral kinetic theory in 
Refs.~\cite{Chen:2015gta,Hidaka:2016yjf,Hidaka:2017auj}.)
For simplicity, we first consider the $\U(1)$ charges, and we will extend our 
results to the $\U(k)$ charges ($k \in {\mathbb N}$) later.

Let us start with the semiclassical action for a chiral fermion with charge $e$ 
in the presence of electromagnetic fields \cite{Son:2012wh,Stephanov:2012ki,Son:2012zy},
\beq
\label{S}
S = \int \left[({\bm p} + e{\bm A}) \cdot {\rm d}{\bm x} - (\epsilon_{\bm p} + e\phi){\rm d}t
-{\bm a}_{\bm p} \cdot {\rm d} {\bm p} \right],
\eeq
where $A^{\mu} =(\phi, {\bm A})$ is the gauge field and ${\bm a}_{\bm p}$ is 
the Berry connection. The Berry curvature is defined from ${\bm a}_{\bm p}$ as 
\beq
\label{Berry}
{\bm \Omega}_{\bm p} \equiv {\bm \nabla}_{\bm p} \times {\bm a}_{\bm p} = K \frac{\hat {\bm p}}{2|{\bm p}|^2}\,,
\eeq
where $\hat {\bm p} \equiv {\bm p}/|{\bm p}|$ is a unit vector and 
$K = \pm 1$ corresponds to right- and left-handed fermions, respectively.
More generically, a Berry monopole with any integer $K$ is possible; 
see, e.g., Refs.~\cite{Xu, Fang} for such realizations in Weyl semimetals. 
For this reason, we consider generic integer $K$ below and will take $K= \pm 1$ later.

The Berry curvature can be regarded as a fictitious magnetic field of a magnetic 
monopole in momentum space \cite{Berry}, whose charge is defined by
\beq
\label{Berry_charge}
\frac{1}{2\pi} \int {\bm \Omega}_{\bm p} \cdot {\rm d}{\bm S} = K\,.
\eeq
This ``Berry monopole" should not be confused with the magnetic monopole
in position space that we will consider later. The chirality of the fermion is taken 
into account by the Berry connection ${\bm a}_{\bm p}$ in the action and by 
the gauge-independent Berry curvature ${\bm \Omega}_{\bm p}$ in the 
following equations of motion and the kinetic theory.

The dispersion relation of chiral fermions also receives a modification by the 
magnetic moment,
\beq
\label{epsilon}
\epsilon_{\bm p} = |{\bm p}| (1 - {\bm \Omega}_{\bm p} \cdot e{\bm B})\,,
\eeq
as required by the Lorentz symmetry of the system \cite{Son:2012zy,Chen:2014cla}.

Here and below, we assume that $e|{\bm E}|, e|{\bm B}| \ll |{\bm p}|^2$. 
We also note that the chiral kinetic theory is applicable to the region 
sufficiently away from the origin ${\bm p} = {\bm 0}$ in momentum space 
\cite{Son:2012wh,Son:2012zy,Stephanov:2012ki}. 
We denote the IR and UV cutoffs of this effective theory as 
$\Delta_{\rm IR, UV} > 0$, respectively, for later purpose.

The equations of motion that follow from the action (\ref{S}) read
\begin{align}
\label{x_dot0}
\dot {\bm x} & = \tilde {\bm v} + \dot {\bm p} \times {\bm \Omega}_{\bm p}, 
\\
\label{p_dot0}
\dot {\bm p} & = e \tilde {\bm E} + \dot {\bm x} \times e{\bm B},
\end{align}
where 
\beq
\tilde {\bm v} = \frac{\d \epsilon_{\bm p}}{\d {\bm p}}\,, \qquad 
e \tilde {\bm E} = e{\bm E} - \frac{\d \epsilon_{\bm p}}{\d {\bm x}}\,.
\eeq
From Eqs.~(\ref{x_dot0}) and (\ref{p_dot0}), we get
\begin{align}
\label{x_dot}
\sqrt{\omega}\dot {\bm x} & = \tilde {\bm v} + e \tilde {\bm E} \times {\bm \Omega}_{\bm p} + 
(\tilde {\bm v} \cdot {\bm \Omega}_{\bm p}) e{\bm B},
\\
\label{p_dot}
\sqrt{\omega}\dot {\bm p} & = e \tilde {\bm E} + \tilde {\bm v} \times e{\bm B} 
+ e^2(\tilde {\bm E} \cdot {\bm B}) {\bm \Omega}_{\bm p},
\end{align}
where $\omega = (1 + e{\bm B} \cdot {\bm \Omega}_{\bm p})^2$.

Now consider the kinetic equation
\beq
\label{Boltzmann}
\frac{\d n_{\bm p}}{\d t} + \dot {\bm x} \cdot \frac{\d n_{\bm p}}{\d {\bm x}}
+ \dot {\bm p} \cdot \frac{\d n_{\bm p}}{\d {\bm p}} = C[n_{\bm p}],
\eeq
where $n_{\bm p}({\bm x})$ is the distribution function of the chiral fermion
in phase space and $C[n_{\bm p}]$ is the collision term, for which we assume
\beq
\int_{\bm p} \sqrt{\omega}C[n_{\bm p}] = 0\,, \qquad
\int_{\bm p} \equiv \int \frac{{\rm d}^3{\bm p}}{(2\pi)^3}\,.
\eeq
The detailed form of $C[n_{\bm p}]$, which can be found in 
Refs.~\cite{Chen:2015gta,Hidaka:2016yjf,Hidaka:2017auj} 
(in a generic reference frame), will be irrelevant in our discussion.
Inserting Eqs.~(\ref{x_dot}) and (\ref{p_dot}) into Eq.~(\ref{Boltzmann}), 
we derive the chiral kinetic equation \cite{Son:2012zy,Manuel:2014dza},
\begin{gather}
\sqrt{\omega} \frac{\d n_{\bm p}}{\d t} + \left[\tilde {\bm v} + e \tilde {\bm E} \times {\bm \Omega}_{\bm p} 
+ (\tilde {\bm v} \cdot {\bm \Omega}_{\bm p})e{\bm B} \right] \cdot \frac{\d n_{\bm p}}{\d {\bm x}}
\nonumber \\ 
+ \left[e \tilde {\bm E} + \tilde {\bm v} \times e{\bm B} 
+ e^2(\tilde {\bm E} \cdot {\bm B}) {\bm \Omega}_{\bm p} \right] \cdot \frac{\d n_{\bm p}}{\d {\bm p}} 
= \sqrt{\omega} C[n_{\bm p}]\,.
\label{CKT}
\end{gather}
One can easily see that, without the Berry curvature, Eq.~(\ref{CKT}) reduces 
to the conventional kinetic equation with electromagnetic fields. 

From Eqs.~(\ref{x_dot}) and (\ref{p_dot}), one can define the fermion number 
and current as \cite{Son:2012wh,Son:2012zy,Stephanov:2012ki}
\begin{align}
\label{n_CKT}
n &\equiv \int_{\bm p} \sqrt{\omega}n_{\bm p}\,, \\
\label{j_CKT}
{\bm j} &\equiv \int_{\bm p} \sqrt{\omega} \dot {\bm x} n_{\bm p}
= \int_{\bm p} \left[\tilde {\bm v} + e\tilde {\bm E} \times {\bm \Omega}_{\bm p} 
+ (\tilde {\bm v} \cdot {\bm \Omega}_{\bm p})e{\bm B} \right] n_{\bm p}\,,
\end{align}
respectively.
The second and third terms in Eq.~(\ref{j_CKT}) are the anomalous Hall effect (AHE)
and the chiral magnetic effect (CME) of the forms 
${\bm j}_{\rm AHE} = e {\bm E} \times {\bm \sigma}_{\rm AHE}$ and
${\bm j}_{\rm CME} = e \sigma_{\rm CME} {\bm B}$ \cite{Vilenkin:1980fu,Nielsen:1983rb,Fukushima:2008xe}, 
respectively, where nonequilibrium expressions of ${\bm \sigma}_{\rm AHE}$ 
and $\sigma_{\rm CME}$ are given by \cite{Son:2012wh,Son:2012zy,Stephanov:2012ki}
\begin{align}
\label{AHE}
{\bm \sigma}_{\rm AHE} &= \int_{\bm p} {\bm \Omega}_{\bm p} n_{\bm p}\,, \\
\label{CME}
\sigma_{\rm CME} &= \int_{\bm p} (\tilde {\bm v} \cdot {\bm \Omega}_{\bm p}) n_{\bm p}
= - \int_{\bm p} \epsilon_{\bm p} {\bm \Omega}_{\bm p} \cdot \frac{\d n_{\bm p}}{\d {\bm p}} \,.
\end{align}
So far, we only consider the contribution of particles, but if antiparticles also 
exist in the system, such contributions have to be added in Eqs.~(\ref{AHE}) 
and (\ref{CME}) as well \cite{Manuel:2013zaa}.

\emph{Dirac monopole in chiral matter.}---%
From now on, we consider the situation where a magnetic monopole in 
position space is present in a many-body system of single right-handed 
(or left-handed) chiral fermions. 

Before considering a generic electromagnetic field configuration, it is 
instructive to consider the case only with the magnetic field of a stationary 
magnetic monopole with charge $g$,
\beq
\label{monopole}
{\bm B}_{\rm mono} = g \frac{\hat {\bm x}}{|\bm x|^2}\,,
\eeq
where we set the monopole at the origin ${\bm x} = {\bm 0}$ without loss 
of generality. In this case, one expects the angular momentum conservation 
due to the rotational symmetry of the system. 

We define ${\bm L} \equiv {\bm x} \times {\bm p}$, and then we have
\begin{align}
\label{Ldot}
\dot {\bm L}&= \dot {\bm x} \times {\bm p} + {\bm x} \times \dot {\bm p} \nonumber \\
&=(\dot {\bm p} \times {\bm \Omega}_{\bm p}) \times {\bm p} + {\bm x} \times (\dot {\bm x} \times e{\bm B}_{\rm mono})
\nonumber \\
&= \frac{\rm d}{{\rm d}t}\left(eg \hat{\bm x} - \frac{K}{2} \hat {\bm p} \right)\,,
\end{align}
where we used the equations of motion~(\ref{x_dot0}) and (\ref{p_dot0}) 
with setting ${\bm E}={\bm 0}$ and ${\bm B} = {\bm B}_{\rm mono}$. 
Equation (\ref{Ldot}) is equivalent to the total angular momentum conservation law:
\beq
\dot {\bm J}={\bm 0}, \qquad {\bm J} \equiv {\bm L} - eg \hat{\bm x} + \frac{K}{2} \hat {\bm p}\,.
\eeq
The second term in ${\bm J}$ corresponds to the angular momentum due to the 
Poynting vector ${\bm E} \times {\bm B}$ in the presence of the Dirac monopole 
in position space. On the other hand, the third term with the coefficient 
$K/2 = \pm 1/2$ corresponds to the spin of chiral fermion, which is incorporated 
by the Berry monopole in momentum space. Because of the quantization of the 
angular momentum, we have 
\beq
\label{Dirac}
eg= \frac{N}{2},
\eeq
with integer $N$. This is the Dirac quantization condition \cite{Dirac:1931kp}.

\emph{Fermion number violation.}---%
We now consider a generic electromagnetic field configuration. 
The modified Gauss's law for magnetism and Amp\`{e}re's law in the presence 
of a magnetic charge $n_{\rm m}$ and magnetic current ${\bm j}_{\rm m}$ are 
\begin{gather}
\label{Gauss}
{\bm \nabla} \cdot {\bm B} = n_{\rm m}, \\
\label{Ampere}
\frac{\d {\bm B}}{\d t} + {\bm \nabla} \times {\bm E} = -{\bm j}_{\rm m}\,,
\end{gather}
respectively. 

Let us study the fermion number (non)conservation of the system.
Generically, for a given system, information of the conservation laws
associated with some symmetries is encoded in the kinetic theory.
In our case, by performing the integration over ${\bm p}$ in Eq.~(\ref{CKT}) 
and using Eqs.~(\ref{Gauss}) and (\ref{Ampere}), one finds
\begin{align}
\label{conservation}
\frac{\d n}{\d t} + {\bm \nabla} \cdot {\bm j} &= - e^2\int_{\bm p} \left({\bm \Omega}_{\bm p} \cdot \frac{\d n_{\bm p}}{\d {\bm p}} \right){\bm E} \cdot {\bm B} 
\nonumber \\
& \quad - e{\bm j}_{\rm m} \cdot \int_{\bm p} {\bm \Omega}_{\bm p} n_{\bm p}
+ e n_{\rm m} \int_{\bm p} (\tilde {\bm v} \cdot {\bm \Omega}_{\bm p})n_{\bm p}\,.
\end{align}
The first term on the right-hand side can be evaluated by using 
${\bm \nabla}_{\bm p} \cdot {\bm \Omega}_{\bm p} = 0$ for ${\bm p} \neq 0$
and the fact that $n_{\bm p} = 1$ for $|{\bm p}| = \Delta_{\rm IR}$
and $n_{\bm p} = 0$ for $|{\bm p}| = \Delta_{\rm UV}$ as \cite{Son:2012wh}
\beq
\label{anomaly}
\frac{e^2}{(2\pi)^3}{\bm E} \cdot {\bm B} \int {\bm \Omega}_{\bm p} \cdot {\rm d}{\bm S}
= \frac{K e^2}{4\pi^2} {\bm E} \cdot {\bm B}\,.
\eeq
This reproduces the triangle anomalies in quantum field theory \cite{Adler,BellJackiw}, 
where the coefficient is topologically quantized by the monopole charge 
in Eq.~(\ref{Berry_charge}).

On the other hand, the second and third terms in Eq.~(\ref{conservation}) are new; 
one can see that these integrals have exactly the same forms as ${\bm \sigma}_{\rm AHE}$ 
and $\sigma_{\rm CME}$ in Eqs.~(\ref{AHE}) and (\ref{CME}), respectively. 

Putting altogether, we arrive at 
\begin{align}
\label{main}
\frac{\d n}{\d t} + {\bm \nabla} \cdot {\bm j} = \frac{K e^2}{4\pi^2} {\bm E} \cdot {\bm B} 
- e {\bm j}_{\rm m} \cdot {\bm \sigma}_{\rm AHE} + e n_{\rm m} \sigma_{\rm CME} \,,
\end{align}
or in a ``Lorentz covariant" form, 
\beq
\label{main_Lorentz}
\d_{\mu} j^{\mu} = -\frac{K e^2}{32\pi^2} \epsilon^{\mu \nu \alpha \beta} F_{\mu \nu} F_{\alpha \beta}
+ e j_{\rm m}^{\mu} \sigma^{\rm chi}_{\mu}\,,
\eeq
where $j^{\mu} \equiv (n, {\bm j})$, $j_{\rm m}^{\mu} \equiv (n_{\rm m}, {\bm j}_{\rm m})$,
and $\sigma^{\rm chi}_{\mu} \equiv (\sigma_{\rm CME}, -{\bm \sigma}_{\rm AHE})$.

As a further demonstration, we consider a single magnetic monopole at the 
origin, $n_{\rm m} = 4\pi g \delta^3({\bm x})$. By performing the spatial integration, 
imposing the boundary condition ${\bm j} \rightarrow {\bm 0}$ at infinity, 
and using the Dirac quantization condition (\ref{Dirac}), we get 
\beq
\label{main_global}
\frac{{\rm d} Q}{{\rm d} t} = \int_{\bm x} \frac{K e^2}{4\pi^2} {\bm E} \cdot {\bm B} 
+ 2 \pi N \sigma_{\rm CME}({\bm 0}) \,,
\eeq
where
\beq
Q \equiv \int_{\bm x} n\,, \qquad \int_{\bm x} \equiv \int {\rm d}^3 {\bm x}\,.
\eeq
Here, we write $\sigma_{\rm CME}({\bm 0})$ to explicitly show that it is the 
value of the chiral magnetic conductivity at the position of the magnetic monopole, 
${\bm x} = {\bm 0}$. Equation (\ref{main_global}) shows that the fermion number 
conservation is violated at the location of the Dirac monopole in chiral matter,
in addition to the triangle anomalies. 
Note that Eqs.~(\ref{main})--(\ref{main_global}) are valid even when the 
system is away from equilibrium, which are part of our main results.

\emph{Local thermal equilibrium.}---% 
When the system is in local thermal equilibrium characterized by a local 
temperature $T$, chemical potential $\mu$ and local fluid velocity 
$u^{\mu} = \gamma(1, {\bm v})$ with $\gamma \equiv (1- {\bm v}^2)^{-1/2}$, 
then the expressions (\ref{main_Lorentz}) and (\ref{main_global}) are 
further simplified. In this case, one finds
\beq
\label{sigma_chi}
\sigma^{\rm chi}_{\mu} = \frac{K \mu}{4\pi^2} u_{\mu}\,.
\eeq
This is easy to check, in particular, in the local rest frame $u^{\mu}=(1,{\bm 0})$ 
in the regime $\mu \gg T$ with a well-defined Fermi surface, 
where $\epsilon_{\bm p} = \mu$. First, the integral in Eq.~(\ref{AHE}) vanishes 
in the local rest frame, since there is no special direction in momentum space, 
and hence, ${\bm \sigma}_{\rm AHE} = {\bm 0}$. Also, repeating the same trick 
as the computation of Eq.~(\ref{anomaly}), 
the integral (\ref{CME}) becomes \cite{Son:2012wh}
\beq
\label{CME2}
\sigma_{\rm CME} = 
\frac{\mu}{(2 \pi)^3} \int {\bm \Omega}_{\bm p} \cdot {\rm d}{\bm S} 
=  \frac{K \mu}{4\pi^2}\,.
\eeq
where the coefficient is again topologically quantized by the monopole charge 
in Eq.~(\ref{Berry_charge}). In fact, the result (\ref{CME2}) is exact 
independently of $\mu$ and $T$: for generic $T$, one can again derive 
Eq.~(\ref{CME2}) by taking into account the contribution of antiparticles \cite{Manuel:2013zaa}. 
By performing the Lorentz boost for $\sigma_{\mu}^{\rm chi} = (\sigma_{\rm CME}, {\bm 0})$, 
we find Eq.~(\ref{sigma_chi}).

Substituting Eq.~(\ref{sigma_chi}) into Eqs.~(\ref{main_Lorentz}) and (\ref{main_global}), 
we have
\beq
\label{main_equilibrium}
\d_{\mu} j^{\mu} = -\frac{K e^2}{32\pi^2} \epsilon^{\mu \nu \alpha \beta} F_{\mu \nu} F_{\alpha \beta}
+ \frac{K e}{4\pi^2} \mu j_{\rm m}^{\mu} u_{\mu}\,,
\eeq
and 
\beq
\label{main_global2}
\frac{{\rm d} Q}{{\rm d} t} = \int_{\bm x} \frac{K e^2}{4\pi^2} {\bm E} \cdot {\bm B} 
+ \frac{N K}{2 \pi} \mu({\bm 0}) \,,
\eeq
The second term in Eq.~(\ref{main_global2}) clearly demonstrates that this 
new fermion number violation is due to the interplay between the Dirac 
monopole in position space, whose charge is characterized by the integer $N$, 
and the Berry monopole in momentum space, whose charge is characterized 
by the integer $K$.

We can extend these results to relativistic matter with Dirac fermions 
at vector chemical potential $\mu_{\rm V} \equiv (\mu_{\rm R} + \mu_{\rm L})/2$ 
and chiral chemical potential $\mu_{\rm A} \equiv (\mu_{\rm R} - \mu_{\rm L})/2$.
By adding or subtracting the right- and left-handed sectors $K = \pm 1$, we get
\begin{align}
\label{vector}
\d_{\mu} j_{\rm V}^{\mu} &= \frac{e \mu_{\rm A}}{2\pi^2} j_{\rm m}^{\mu} u_{\mu} \,,
\\
\label{axial}
\d_{\mu} j_{\rm A}^{\mu} &= -\frac{e^2}{16\pi^2} \epsilon^{\mu \nu \alpha \beta} F_{\mu \nu} F_{\alpha \beta}
+  \frac{e \mu_{\rm V}}{2\pi^2} j_{\rm m}^{\mu} u_{\mu} \,,
\end{align}
where $j^{\mu}_{\rm V} \equiv j^{\mu}_{\rm R} + j^{\mu}_{\rm L}$ and 
$j^{\mu}_{\rm A} \equiv j^{\mu}_{\rm R} - j^{\mu}_{\rm L}$ are the vector 
and axial-vector currents, respectively.

\emph{Extension to generic charges.}---%
We can also extend the result (\ref{main_equilibrium}) to the case with more 
generic charges. Consider a conserved current 
$j^{\mu a} = \bar \psi \sigma^{\mu} T^a \psi$, where $\psi$ is a chiral fermion field 
of interest, $T^a$  ($a=0,1,\cdots,k^2-1$) are the $\U(k)$ generators, and 
$\sigma^{\mu} = (1, {\bm \sigma})$. We denote the field strength of the $\U(k)$ 
gauge field as $F_{\mu \nu}^a$. In the presence of a four magnetic current 
$j_{\rm m}^{\mu a} = (n_{\rm m}^a, {\bm j}_{\rm m}^a)$, we have
\beq
\label{non-Abelian0}
\d_{\mu} j^{\mu a} = -\frac{Ke^2}{32\pi^2} d^{abc} \epsilon^{\mu \nu \alpha \beta} F_{\mu \nu}^b F_{\alpha \beta}^c
+ d^{abc} e j_{\rm m}^{\mu b} \sigma^{\rm chi}_{\mu c}\,, \nonumber \\
\eeq
in nonequilibrium state. Here, 
$d^{abc} = \frac{1}{2}{\rm tr}(T^a \{T^b, T^c \})$ and
$\sigma^{\rm chi}_{\mu a} \equiv (\sigma_a^{\rm CME}, -{\bm \sigma}_a^{\rm AHE})$
with \cite{Akamatsu:2013pjd,Akamatsu:2014yza}
\begin{align}
\sigma_a^{\rm CME} = - \int_{\bm p} \epsilon_{\bm p} {\bm \Omega}_{\bm p} \cdot \frac{\d n^a_{\bm p}}{\d {\bm p}} \,, \qquad
{\bm \sigma}_a^{\rm AHE} = \int_{\bm p} {\bm \Omega}_{\bm p} n^a_{\bm p}\,, 
\end{align}
and $n^a_{\bm p}$ the distribution function for the $\U(k)$ charge.
In equilibrium, Eq.~(\ref{non-Abelian0}) reduces to
\begin{align}
\label{non-Abelian}
\d_{\mu} j^{\mu a} = 
-\frac{Ke^2}{32\pi^2} d^{abc} \epsilon^{\mu \nu \alpha \beta} F_{\mu \nu}^b F_{\alpha \beta}^c
+ \frac{Ke}{4\pi^2} d^{abc} \mu^b j_{\rm m}^{\mu c} u_{\mu} \,,
\end{align}
where $\mu^a$ is the chemical potential for the conserved current $j^{\mu a}$.

\emph{Relation to the Witten effect.}---% 
We note in passing that Eq.~(\ref{vector}) has a connection with 
the Witten effect \cite{Witten:1979ey}. 
To make this connection clear, we again consider the spatial integral of 
Eq.~(\ref{vector}), which yields
\beq
\label{Q_V}
\frac{{\rm d} Q_{\rm V}}{{\rm d} t} = \frac{N}{\pi} \mu_{\rm A}({\bm 0})\,,
\qquad Q_{\rm V} \equiv \int_{\bm x} n_{\rm V}\,.
\eeq

We now recall the similarity of the physics at finite $\mu_{\rm A}$ to 
the physics at finite $\theta$ angle \cite{Kharzeev:2009fn}, 
according to which $2 \mu_{\rm A}$ corresponds to $\dot \theta$. 
By making the replacement $\mu_{\rm A} \rightarrow \dot \theta/2$ in 
Eq.~(\ref{Q_V}) and performing the integration under the condition that 
the additional vector charge due to the monopole vanishes 
($\Delta Q_{\rm V}=0$) when $\theta = 0$, we obtain
\beq
\Delta Q_{\rm V} = \frac{N}{2\pi} \theta\,.
\eeq
This is exactly the Witten effect \cite{Witten:1979ey}.
Note however that our results above cannot simply be retrieved from the 
Witten effect itself. Note also that such a connection would be lost in 
nonequilibrium.

\emph{Discussions.}---%
In this paper, we derived the model-independent expressions of the anomalous 
fermion number violation due to the presence of the Dirac monopole in 
equilibrium and nonequilibrium chiral matter. Our main results are Eqs.~(\ref{main_Lorentz}), 
(\ref{main_global}) and (\ref{main_equilibrium})--(\ref{non-Abelian}). 

Our argument so far is general in that it does not rely on the details of systems. 
Let us now consider a grand unified theory (GUT) as an example, where 
$\U(1)$ electromagnetism is embedded in a ``spontaneously broken gauge symmetry," 
leading to a magnetic monopole as a topological soliton. If such a GUT are chiral, 
a finite chiral asymmetry of a fermion can be produced at the GUT scale 
(see, e.g., Refs.~\cite{Joyce:1997uy,Kamada:2018tcs}).
Then, our results show that the finite chiral conductivity $\sigma^{\rm chi}_{\mu a}$ 
gives a new contribution to the baryon and/or lepton number violation in addition 
to the quantum anomalies. For a given model, our formulation and mechanism 
should allow one to compute such violations quantitatively. This may be relevant 
to the nonequilibrium evolution of the early Universe including the problems of 
baryogenesis and leptogenesis.

\acknowledgments
This work was supported by the Keio Institute of Pure and Applied Sciences (KiPAS) project at Keio University
and JSPS KAKENHI grant No.~19K03852.

\end{document}